\documentclass[showpacs,pra,preprint]{revtex4-1}
\usepackage{graphicx}
\usepackage{dcolumn}
\usepackage[latin5]{inputenc}
\usepackage{amssymb,amsmath,mathptmx,amsbsy,bm}
\def\bra#1{\mathinner{\langle{#1}|}}
\def\ket#1{\mathinner{|{#1}\rangle}}
\def\braket#1{\mathinner{\langle{#1}\rangle}}

\begin{document}
\title{Optimal Distillation of Three Qubit $W$ States}
\author{Ali Yildiz}
\affiliation{Department of Physics, Istanbul Technical University, Maslak 34469, Istanbul, Turkey}
\date{\today}
\email{yildizali2@itu.edu.tr}

\begin{abstract}

Some of the asymmetric  three qubit $W$  states are used for perfect
teleportation, superdense coding and quantum information splitting. We present the  protocols for the optimal distillation of the asymmetric as well as the symmetric $W$ states   from a single copy of any three qubit $W$ class pure state.
\end{abstract}
 \maketitle
\section{Introduction}
The use of entanglement as a resource  in quantum information and quantum computation requires characterization, manipulation and quantification problems to be solved. The bipartite pure state entanglement has been well understood. In the two qubit case Einstein-Podolski-Rosen  (EPR) states $\frac{1}{\sqrt{2}}(\ket{00}+\ket{11})$  are used in many quantum information processes  such as perfect teleportation \cite{teleportation1} and dense coding \cite{densecoding}.
If the initial state used as a resource is not an EPR state but any state in the canonical form $a\ket{00}+b\ket{11}$  $(a\geq b\geq 0)$ then it is possible to perform the task  with a maximum probability of success $2b^2$ \cite{ProbTELEP1,ProbTELEP2,ProbDENSE1}. An alternative way is distilling  \cite{dist} the EPR state by performing the positive operator valued measurement (POVM) $ \frac{b}{a}\ket{0}\bra{0}+\ket{1}\bra{1}$ on the first qubit and then perform the teleportation or dense coding with unit probability. The success probability of the distilling an EPR state turns out to be $2b^2$.

The many body entanglement is not a straight forward generalization of the bipartite case and some challenging problems still remain unsolved.  In the three qubit case, for example, three are two classes of tripartite entangled states which can not be converted into each other by stochastic local operations and classical communication (SLOCC) \cite{GHZ_W_inequiv}, namely the GHZ  and  $W$ class states. Any two states of the same class can be converted into each other by means of SLOCC.  The GHZ state $\ket{GHZ}=\frac{1}{\sqrt{2}}(\ket{000}+\ket{111})$ and the symmetric $W$ state
\begin{equation}\label{W}
\ket{W}=\frac{1}{\sqrt{3}}(\ket{001}+\ket{010}+\ket{100})
\end{equation}
are considered as the representatives of the GHZ and $W$ classes respectively.
Although the symmetric  $W$ state (\ref{W}) is more robust against decoherence or particle loses it can not be used to perform  perfect quantum information tasks  \cite{teleportation2,teleportation3}. The asymmetric $W$  states
\begin{equation}\label{Agrawal1}
\frac{1}{\sqrt{2}}\ket{001}+\frac{1}{2}\ket{010}+\frac{1}{2}\ket{100},
\end{equation}
\begin{equation}\label{zheng}
\frac{1}{2}\ket{001}+\frac{1}{2}\ket{010}+\frac{1}{\sqrt{2}}\ket{100}
\end{equation}
are widely used in perfect quantum information processes \cite{W1,W2,W3,W4}. If the three qubit entangled state which will be used as a resource is not a GHZ state or  an asymmetric $W$  state (\ref{Agrawal1})-(\ref{zheng}) then the distillation these states is necessary to successfully perform quantum information tasks.
The optimal distillation  of the GHZ state from a single copy of GHZ class state  is presented in \cite{Optimal_GHZ_distil}. Although   the distillation of the symmetric $W$ state (\ref{W}) from some special $W$ states is presented in \cite{Wdistil1,Wdistil2} the  protocol for the optimal distillation of the asymmetric $W$ states is still unknown. In this work we present the protocol for the optimal distillation of the asymmetric $W$ states (\ref{Agrawal1})-(\ref{zheng}) as well as the symmetric $W$ state (\ref{W}) from an arbitrary $W$ class state.
The procedure we use is as follows: we first define the canonical form of the $W$ class states   which any $W$ class state can be brought into by local unitary transformations and then show that the general local POVMs  followed by local unitary transformations which bring the state into canonical form equal to the local POVMs  which leave the canonical form invariant.  We then find the local POVMs to maximize the probability of obtaining symmetric or asymmetric   $W$ states (\ref{W})-(\ref{zheng}).
\section{Canonical Form of W Class States and General Local POVMs }

To define the canonical form of the $W$ class states which we are going to use in the distillation let us first review the canonical form of any three qubit pure state as defined in \cite{cano-form1,cano-form2}: any three qubit state
\begin{equation}
\ket{\psi}=\sum_{ijk}{t_{ijk}\ket{ijk}}
\end{equation}
 defines matrices  $T_0$ and $T_1$ by

\begin{equation}
\ket{\psi}=\sum_{jk}{T_{0,jk}\ket{0}\ket{jk}+T_{1,jk}\ket{1}\ket{jk}}.
\nonumber
\end{equation}
Under the unitary transformations on the first qubit   the matrices $T_0$ and $T_1$ transform as
\begin{eqnarray}
T_0'&=&u_{00}^AT_0+u_{01}^AT_1 \nonumber\\
T_1'&=&u_{10}^AT_0+u_{11}^AT_1 \quad , \quad
u_{tz}=\braket{t|U|z}.
\end{eqnarray}
It is always possible to make $detT_0'=0$ and the unitary transformations on the second and third qubits diagonalize $T_0'$. Then the canonical form of the generic three qubit states is defined  by
\begin{equation}\label{canonical}
\ket{\psi}  =  \lambda_0\ket{000}+\lambda_1e^{i\varphi}\ket{100}+\lambda_2\ket{101}+ \lambda_3\ket{110}+\lambda_4\ket{111},
\quad \lambda_i \geq 0.
\end{equation}
There are two solutions for $detT_0'=0$ and hence there are two sets of values of $\lambda$s for any generic state. If the 3-tangle \cite{three-tangle} given by $\lambda_0\lambda_4$ is nonzero then the three qubit state is of GHZ class. If 3-tangle is zero and  the rank of the  reduced density matrices  $\rho_A\equiv$Tr$_{BC}\ket{\psi}\bra{\psi}$, $\rho_B$ and $\rho_C$ are two
then the states are of $W$ class states. Without loss of generality $\lambda_4=0$ and it turns out that  there is only one set of $\lambda$s in (\ref{canonical}) for $W$ class states. We also  use the fact that the permutation of the parties $1\leftrightarrow2$ gives $\lambda_0\leftrightarrow\lambda_3$,  $1\leftrightarrow3$ gives $\lambda_0\leftrightarrow\lambda_2$ and $2\leftrightarrow3$ gives $\lambda_2\leftrightarrow\lambda_3$  and define the canonical form of any three qubit $W$ class state  as

\begin{equation}\label{k}
\ket{\psi}=\lambda_0\ket{000}+\lambda_1\ket{100}+\lambda_2\ket{101}+\lambda_3\ket{110}\quad (\lambda_0\geq\lambda_2\geq\lambda_3).
\end{equation}
We also note that $\lambda_1$ is invariant under the permutation of the parties and the two asymmetric $W$ states (\ref{Agrawal1}) and (\ref{zheng}) are equal up to the permutation of the parties $1\leftrightarrow3$ and hence their canonical forms are equal and given by
\begin{equation}\label{asymW}
\frac{1}{\sqrt{2}}\ket{000}+\frac{1}{2}\ket{100}+\frac{1}{2}\ket{101}.
\end{equation}
The canonical form of the symmetric $W$  state (\ref{W}) is also found to be
\begin{equation}\label{symW}
\ket{W}=\frac{1}{\sqrt{3}}(\ket{000}+\ket{100}+\ket{101}).
\end{equation}
We now consider that a general local POVM
\begin{equation}
A'=e^{i\theta_{1}}a\ket{0}\bra{0}+e^{i\theta_{2}}b\ket{0}\bra{1}+e^{i\theta_{3}}c\ket{1}\bra{0}+e^{i\theta_{4}}d\ket{1}\bra{1}\quad (a, b, c, d \  \textnormal{real})
\end{equation}
is performed on the first qubit which transforms the state (\ref{k}) into
\begin{equation}
\ket{\psi'}=\frac{1}{\sqrt{p_A}}(A'\otimes I_B\otimes I_C)\ket{\psi}
\end{equation}
with probability $p_A=(\bra{\psi}(A')^{\dag}A'\otimes I_B\otimes I_C)\ket{\psi}$ and then the resulting state is brought into the canonical form by local unitary transformations to give

\begin{eqnarray}\label{oper1}
& &\ket{\psi'}= \frac{1}{\sqrt{p_A}}( \lambda_0\frac{\left|e^{i(\theta_{1}+\theta_{4})}ad-e^{i(\theta_{2}+\theta_{3})}bc\right|}{\sqrt{b^2+d^2}}\ket{000}+ \\
& &\left|\lambda_0\frac{e^{i(\theta_{1}-\theta_{2})}ab+e^{i(\theta_{3}-\theta_{4})}cd}{\sqrt{b^2+d^2}}+\lambda_1\sqrt{b^2+d^2}\right|\ket{100}+\lambda_2\sqrt{b^2+d^2}\ket{101}+\lambda_3\sqrt{b^2+d^2}\ket{110})\nonumber.
\end{eqnarray}
Using the fact that the POVM
\begin{eqnarray}
& &A=\frac{\left|e^{i(\theta_{1}+\theta_{4})}ad-e^{i(\theta_{2}+\theta_{3})}bc\right|}{\sqrt{b^2+d^2}}\ket{0}\bra{0}+\\
& &(\left|\frac{e^{i(\theta_{1}-\theta_{2})}ab+e^{i(\theta_{3}-\theta_{4})}cd}{\sqrt{b^2+d^2}}+\frac{\lambda_1\sqrt{b^2+d^2}}{\lambda_0}\right|-\frac{\lambda_1\sqrt{b^2+d^2}}{\lambda_0})\ket{1}\bra{0}+\sqrt{b^2+d^2}\ket{1}\bra{1}\nonumber
\end{eqnarray}
on the first qubit transforms the state (\ref{k}) into state  (\ref{oper1}) with the same probability $p_A$ we conclude that the most general POVM on the first qubit is of the form
\begin{equation}
A=a_1\ket{0}\bra{0}+c_1\ket{1}\bra{0}+d_1\ket{1}\bra{1}\quad (a_1, c_1, d_1\  \textnormal{real}) .
\end{equation}
It can similarly be shown that the general local POVMs on the second and third qubits are of the form
\begin{eqnarray}
B&=&a_2\ket{0}\bra{0}+b_2\ket{0}\bra{1}+d_2\ket{1}\bra{1}\quad (a_2, b_2, d_2\  \textnormal{real})  \\
C&=&a_3\ket{0}\bra{0}+b_3\ket{0}\bra{1}+d_3\ket{1}\bra{1}\quad (a_3, b_3, d_3\  \textnormal{real}) \nonumber.
\end{eqnarray}
The condition that eigenvalues of $A^{\dagger}A$,  $B^{\dagger}B$ and $C^{\dagger}C$ should be less than or equal to one gives the constraints
\begin{eqnarray}\label{constraints}
& &a_i^2+b_i^2+d_i^2+\sqrt{((a_i - d_i)^2 + b_i^2) ((a_i + d_i)^2 + b_i^2)} \leq 2,\quad i=2, 3\nonumber \\
& &a_1^2+c_1^2+d_1^2+\sqrt{((a_1 - d_1)^2 + c_1^2) ((a_1 + d_1)^2 + c_1^2)} \leq 2.
\end{eqnarray}
Hence the most general transformation of the state ($\ref{k}$) under local POVMs is given by
\begin{eqnarray}\label{kano}
\ket{\psi'}&=&\frac{1}{\sqrt{P}}A\otimes B \otimes C\ket{\psi}\nonumber \\
&=&\frac{1}{\sqrt{P}}(\lambda_0a_1a_2a_3\ket{000}+((\lambda_0c_1+\lambda_1d_1)a_2a_3+
\lambda_2d_1a_2b_3+\lambda_3d_1b_2a_3)\ket{100}\nonumber \\ & &+\lambda_2d_1a_2d_3\ket{101}+\lambda_3d_1d_2a_3\ket{110})
\end{eqnarray}
where $P=\bra{\psi}A^{\dag}A\otimes B^{\dag}B\otimes C^{\dag}C\ket{\psi}$.

\section{Optimal Distillation of the Asymmetric W States}
We now discuss the optimal distillation of the asymmetric $W$ state (\ref{asymW}): for
\begin{eqnarray}\label{cond}
\lambda_0a_1a_2a_3=\sqrt{2} \lambda_2d_1a_2d_3=\sqrt{2} \lambda_3d_1d_2a_3,
 (\lambda_0c_1+\lambda_1d_1)a_2a_3+
\lambda_2d_1a_2b_3+\lambda_3d_1b_2a_3=0
\end{eqnarray}
the resulting state is (\ref{asymW}) and the probability of success turns out to be
\begin{equation}\label{prob}
P=2\lambda_0^2a_1^2 a_2^2 a_3^2.
\end{equation}
Now the problem is to find the local POVMs to maximize the probability. The maximization of the local probabilities
\begin{equation}\label{OSBP}
det(I_A -A^{\dag} A)=0,\ det(I_B-B^{\dag} B)=0,\ det(I_C -C^{\dag} C)=0
\end{equation}
implies that the constraints
\begin{eqnarray}\label{max2}
(1-a_1^2)(1-d_1^2)&=& c_1^2,\nonumber \\
(1-a_2^2)(1-d_2^2) &=& b_2^2, \\
 (1-a_3^2)(1-d_3^2)&=& b_3^2\nonumber
\end{eqnarray}
should be satisfied and  the state is either transformed into (\ref{asymW}) or otherwise disentangled, i.e., we are using \emph{one successful branch protocol} (OSBP). The problem of optimal distillation of the state (\ref{asymW}) using OSBP is reduced to the problem of maximizing  $(\ref{prob})$ subject to the constraints (\ref{constraints}), (\ref{cond}) and (\ref{max2}). Defining $y\equiv a_3^2$ the maximum probability is found to be the maximum of the function
\begin{equation}\label{solution1}
P(y)=\lambda _2^2+\lambda _0^2y+\lambda _1^2y-\lambda _2^2y+\lambda _3^2y+\lambda _3^2y^2-2K-\sqrt{L+M},\ \ (0<y\leq 1)
\end{equation}
where
\begin{eqnarray}\label{defs}
K&=&\lambda _1\sqrt{y(1-y)(\lambda _2-\lambda _3^2y)},\nonumber \\
L&=&\lambda _2^4(1-y)^2+\lambda _0^4y^2+\lambda _1^4y^2-2\lambda _1^2\lambda _3^2y^2+\lambda _3^4y^2+6\lambda _1^2\lambda _3^2y^3+2\lambda _3^4y^3\nonumber\\
 & &+\lambda _3^4y^4-4y(\lambda _1^2-\lambda _3^2-\lambda _3^2y)K,\\
M&=&2\lambda _2^2(1-y)(\lambda _0^2y+3\lambda _1^2y+\lambda _3^2y+\lambda _3^2y^2-2K)+2\lambda _0^2(\lambda _1^2y^2+\lambda _3^2(y-3)y^2-2yK).\nonumber
\end{eqnarray}
The solutions to the local POVMs are given by

\begin{eqnarray}\label{sol2}
a_1=\sqrt{\frac{P}{2\lambda_0^2a_3^2}},\quad d_1&=&\frac {\lambda_0}{\sqrt{2} \lambda_3}a_1,\quad c_1=\sqrt{(1-a_1^2)(1-\frac {2 \lambda_3^2}{\lambda_0^2}a_1^2)},\nonumber\\
a_2=1,\quad b_2&=&0,\quad d_2=1\\
d_3=\frac {\lambda_3}{\lambda_2}a_3,& &\quad b_3=\sqrt{(1-a_3^2)(1-\frac {\lambda_3^2}{\lambda_2^2}a_3^2)}\nonumber
\end{eqnarray}
where $P$ given by (\ref{solution1}) is a function of $a_3$.
We note that the case y=0 ($a_3=0$) implies that the rank of the operator C is one, i.e, the third party makes  projective measurement which  disentangles the third particle from the other two.

To prove that no distillation protocol can give a greater probability one needs to show that the inequality
\begin{equation}\label{mono1}
 P(\ket{\psi})\geq\sum_{i}p_iP(\ket{\psi_i})
\end{equation}
is satisfied for any sequence of local quantum operations that transform $\ket{\psi}$ into $\ket{\psi_i}$ with probability $p_i$. The right hand side of the inequality (\ref{mono1}) is the average probability to obtain the state (\ref{asymW}) using several branches whereas left hand side is the probability for OSBP. Taking into account that any POVM can be decomposed into a sequence of two-outcome POVMs \cite{Optimal_GHZ_distil} it is sufficient to show
\begin{equation}\label{mono2}
P(\ket{\psi})\geq p_1P(\ket{\psi_1})+p_2P(\ket{\psi_2})
\end{equation}
where $\ket{\psi_1}$ and $\ket{\psi_2}$ are obtained by the most general POVMs on one of the qubits, say the first qubit. We start with a two outcome POVM with operators
\begin{eqnarray}\label{oper2}
A_1&=&a_1\ket{0}\bra{0}+c_1\ket{1}\bra{0}+d_1\ket{1}\bra{1},\nonumber \\ A_2&=&\alpha_1\ket{0}\bra{0}+\gamma_1\ket{1}\bra{0}+\delta_1\ket{1}\bra{1}
\end{eqnarray}
acting on the first qubit satisfying $A_1^{\dag}A_1+A_2^{\dag}A_2=I$. The states
\begin{eqnarray}\label{stat3}
\ket{\psi_1} &= &\frac{1}{\sqrt{p_1}}(\lambda_0a_1\ket{000}+(\lambda_0c_1+\lambda_1d_1)\ket{100}+\lambda_2d_1\ket{101}+\lambda_3d_1\ket{110}),\nonumber \\
\ket{\psi_2} &=& \frac{1}{\sqrt{p_2}}(\lambda_0f_1\ket{000}+(\lambda_0g_1+\lambda_1h_1)\ket{100}+\lambda_2h_1\ket{101}+\lambda_3h_1\ket{110})
\end{eqnarray}
are obtained with probabilities $p_i=\bra{\psi}A_i^{\dag}A_i\ket{\psi}$. Then OSBP is used on the states $\ket{\psi_1}$ and $\ket{\psi_2}$ to give the maximum probabilities  $P(\ket{\psi_1})$ and $P(\ket{\psi_2})$ for the distillation  of the asymmetric $W$ state (\ref{asymW}).  To check if the inequality (\ref{mono2}) is satisfied we maximize   $p_1P(\ket{\psi_1})+p_2P(\ket{\psi_2})$ and find that the maximum is obtained for  $P(\ket{\psi_1})=0$ or $P(\ket{\psi_2})=0$ which means that  no distillation protocol can produce a higher probability of success than the OSBP  we present.

The maximization of (\ref{solution1}) requires numerical calculations in general. For illustrative purposes we discuss the special case  $\lambda_1=0$, i.e.,
the optimal distillation of the state (\ref{asymW}) using the $W$ class state
\begin{equation}\label{special1}
\ket{\psi}=\lambda_0\ket{000}+\lambda_2\ket{101}+\lambda_3\ket{110}\quad (\lambda_0\geq\lambda_2\geq\lambda_3).
\end{equation}
For $\lambda_1=0$ the maximum of the probability function (\ref{solution1}) is given by
\begin{equation}\label{solspecial}
P=\lambda_0^2+2\lambda_3^2-\mid \lambda_0^2-2\lambda_3^2\mid
\end{equation}
at the point $y=1$ ($a_3=1$) and the local POVMs turn out to be
\begin{eqnarray}\label{special12}
A=\frac{\sqrt{2}\lambda_3}{\lambda_0}\ket{0}\bra{0}+\ket{1}\bra{1}\quad (\sqrt{2}\lambda_3\leq \lambda_0),& & \quad A=\ket{0}\bra{0}+\frac{\lambda_0}{\sqrt{2}\lambda_3}\ket{1}\bra{1}\quad (\sqrt{2}\lambda_3\geq \lambda_0),\nonumber \\
B=\ket{0}\bra{0}+\ket{1}\bra{1},& &\quad
C=\ket{0}\bra{0}+\frac{\lambda_3}{\lambda_2}\ket{1}\bra{1}.
\end{eqnarray}
As an immediate application we consider the teleportation using the symmetric $W$ state as a resource. It is possible to perform the teleportation with unit fidelity but with success probability 2/3 \cite{Wtele} which means that the probability of losing the information is 1/3. However if  we  prefer not to lose the information to be teleported, we first distill the asymmetric state (\ref{asymW})  by the local operations
\begin{eqnarray}\label{symWdist}
\quad A=\ket{0}\bra{0}+\frac{1}{\sqrt{2}}\ket{1}\bra{1},\quad
B=\ket{0}\bra{0}+\ket{1}\bra{1},\quad
C=\ket{0}\bra{0}+\ket{1}\bra{1}
\end{eqnarray}
with success probability 2/3 and then o perform the teleportation \cite{ProbTELEP2} with unit fidelity and unit success probability.  If the distillation is not successful we keep the state to be teleported.
\section{Optimal Distillation of the Symmetric W States}
Our method can also be used  for the optimal distillation of the symmetric $W$  state (\ref{symW}) from an arbitrary $W$ class state (\ref{k}). We again consider the most general local transformations of the $W$ class state (\ref{kano}) and  find the local operations which maximize the probability of obtaining the symmetric $W$  state (\ref{symW}). For
\begin{eqnarray}\label{cond2s}
\lambda_0a_1a_2a_3= \lambda_2d_1a_2d_3= \lambda_3d_1d_2a_3,
 (\lambda_0c_1+\lambda_1d_1)a_2a_3+
\lambda_2d_1a_2b_3+\lambda_3d_1b_2a_3=0
\end{eqnarray}
the resulting state is a symmetric $W$  state (\ref{symW}) with the probability of success
\begin{equation}\label{prob2s}
P=3\lambda_0^2a_1^2 a_2^2 a_3^2.
\end{equation}
We now maximize (\ref{prob2s}) under the constraints (\ref{constraints}), (\ref{max2}) and (\ref{cond2s}). Defining $y\equiv a_3^2$ the maximum probability is found to be the maximum of the function
\begin{equation}\label{solution11}
P(y)=\frac{3}{2}(\lambda _2^2+\lambda _0^2y+\lambda _1^2y-\lambda _2^2y+\lambda _3^2y^2-2Q-\sqrt{R+S}),\ \ (0<y\leq 1)
\end{equation}
where
\begin{eqnarray}\label{defs2}
Q&=&\lambda _1\sqrt{y(1-y)(\lambda _2-\lambda _3^2y)},\nonumber \\
R&=&\lambda _2^4(1-y)^2+\lambda _0^4y^2+\lambda _1^4y^2-4\lambda _1^2\lambda _3^2y^2+6\lambda _1^2\lambda _3^2y^3+\lambda _3^4y^4-4\lambda _1^2yQ+4\lambda _3^2y^2Q, \\
S&=&2\lambda _2^2(1-y)(\lambda _0^2y+3\lambda _1^2y+\lambda _3^2y^2-2Q)+\lambda _0^2(2\lambda _1^2y^2+2\lambda _3^2(y-2)y^2-4yQ).\nonumber
\end{eqnarray}
and the solutions to the local POVMs are given by

\begin{eqnarray}\label{sol11}
a_1=\sqrt{\frac{P}{3\lambda_0^2a_3^2}},\quad d_1&=&\frac {\lambda_0}{\lambda_3}a_1,\quad c_1=\sqrt{(1-a_1^2)(1-\frac {\lambda_3^2}{\lambda_0^2}a_1^2)},\nonumber\\
a_2=1,\quad b_2&=&0,\quad d_2=1\\
d_3=\frac {\lambda_3}{\lambda_2}a_3,& &\quad b_3=\sqrt{(1-a_3^2)(1-\frac {\lambda_3^2}{\lambda_2^2}a_3^2)}\nonumber
\end{eqnarray}
where $P$ given by (\ref{solution11}) is a function of $a_3$. To show that the POVMs given by (\ref{sol11}) gives the optimal distillation protocol  we need to show that the inequality (\ref{mono2}) is not violated by any two outcome POVM with the operators (\ref{oper2}) and the transformed states (\ref{stat3}). We find that the maximum of the right hand side of the inequality (\ref{mono2}) is obtained for $P(\ket{\psi_1})=0$ or $P(\ket{\psi_2})=0$. This proves that  no distillation protocol can produce a higher probability of success than the OSBP given by (\ref{sol11}). For illustrative purposes we again consider the special case   $\lambda_1=0$, i.e.,
the optimal distillation of the state (\ref{symW}) from the $W$ class state (\ref{special1}). We find that the maximum of the probability function (\ref{solution11}) is $3\lambda_3^2$ at the point $y=1$ ($a_3=1$) and the local POVMs are given by

\begin{equation}\label{special13}
A=\frac{\lambda_3}{\lambda_0}\ket{0}\bra{0}+\ket{1}\bra{1}, \quad
B=\ket{0}\bra{0}+\ket{1}\bra{1},\quad
C=\ket{0}\bra{0}+\frac{\lambda_3}{\lambda_2}\ket{1}\bra{1}.
\end{equation}

\section{Conclusion}
In this work we explicitly constructed the optimal local protocol for the distillation of the symmetric $W$ state (\ref{symW}) as well as the asymmetric $W$ state (\ref{asymW}) which is used in perfect teleportation, dense coding and information splitting.
We note that in contrast to the distillation of  the GHZ states \cite{Optimal_GHZ_distil} where all three parties should perform local POVMs   only two parties should apply POVMs in the distillation of the $W$ states. This is related to the fact that in a general GHZ class state given by (\ref{canonical}) three coefficients ($\lambda_1$, $\lambda_2$ and $\lambda_3$) should be made zero which requires the cooperation of all three parties. Since $\lambda_4$ is zero for $W$ class states it is possible to distill symmetric or asymmetric  $W$ states by the cooperation of only two parties. We have shown that the use of the canonical form and the local POVMs which leave the canonical form invariant simplifies the problem of manipulation of pure states. This result can  be used as an alternative approach for the distillation of the  GHZ states and the manipulation of other many partite pure states.


\begin{thebibliography} {99}

\bibitem{teleportation1} C. H. Bennett, G. Brassard, C. Crepeau, R. Jozsa, A. Peres and W. K. Wootters, Phys. Rev. Lett. {\bf70,} 1895 (1993).

\bibitem{densecoding} C. H. Bennett, and S. J. Wiesner, Phys. Rev. Lett. {\bf69,} 2881 (1992).

\bibitem{ProbTELEP1} W. L. Li, C. F. Li and G. C. Guo, Phys. Rev. A {\bf61,} 034301 (2000).

\bibitem{ProbTELEP2} P. Agrawal and A. K. Pati, Phys. Lett. A {\bf305,} 12 (2002).

\bibitem{ProbDENSE1} J. C. Hao, C. F. Li and G. C. Guo, Phys. Lett. A {\bf278,} 113 (2000).

\bibitem{dist} C. H. Bennett, H. J. Bernstein, S. Popescu and B. Schumacher, Phys. Rev. A {\bf53,} 2046 (1996).

\bibitem{GHZ_W_inequiv} W. Dur, G. Vidal and J. I. Cirac, Phys. Rev. A {\bf62,} 062314 (2000).

\bibitem{teleportation2} J. Joo, Y. J. Park, S. Oh and J. Kim, New J. Phys. {\bf5,} 136 (2003).

\bibitem{teleportation3} V. N. Gorbachev, A. I. Trubilko, A. A. Rodichkina and A. I. Zhiliba, Phys. Lett. A {\bf314,} 267 (2003).

\bibitem{W1} S. B. Zheng, Phys. Rev. A  {\bf74,} 054303 (2006).

\bibitem{W2} P. Agrawal and A. Pati, Phys. Rev. A {\bf74,} 062320 (2006).

\bibitem{W3} L. Li and D. Qiu, J. Phys. A: Math. Theor. {\bf40,} 10871 (2007).

\bibitem{W4} H. Situ and D. Qiu, J. Phys. A: Math. Theor. {\bf43,} 055301 (2010).

\bibitem{Optimal_GHZ_distil} A. Acin, E. Jane, W. Dur and G. Vidal, Phys. Rev. Lett. {\bf85,}  4811 (2000).

\bibitem{Wdistil1} Z. L. Cao and M. Yang, J. Phys. B: At. Mol. Opt. Phys. {\bf36,} 4245 (2003).

\bibitem{Wdistil2} M. Yang and Z. L. Cao, Physica A, {\bf337,} 141 (2004).

\bibitem{cano-form1} A. Acin, A. Andrianov, L. Costa, E. Jane, J. I. Latorre, and R. Tarrach, Phys. Rev. Lett. {\bf85,} 1560 (2000).

\bibitem{cano-form2} A. Acin, A. Andrianov, E. Jane and R. Tarrach,  J. Phys. A: Math. Gen. {\bf34,} 6725 (2001).


\bibitem{three-tangle} V. Coffman, J. Kundu and W. K. Wootters,
Phys. Rev. A {\bf61,} 052306 (2000).

\bibitem{Wtele} B. S. Shi and A. Tomita, Phys. Lett. A {\bf296,} 161 (2002); J. Joo and Y. J. Park, Phys. Lett. A {\bf300,} 324 (2002).



\end{thebibliography}
\end{document}